\documentclass[aps,prd,10pt,twocolumn,floatfix,superscriptaddress,preprintnumbers,shownopacs,showkeys,nofootinbib,notoccite,notitlepage]{revtex4-1}

\usepackage{graphicx}
\usepackage{dcolumn}
\usepackage{footmisc}

\usepackage{amsmath}
\usepackage{bm}
\usepackage{hyperref}
\usepackage{xcolor}
\usepackage{float}
\begin{document}
\setlength{\abovedisplayskip}{5pt}
\setlength{\belowdisplayskip}{5pt}
\setlength{\abovedisplayshortskip}{5pt}
\setlength{\belowdisplayshortskip}{5pt}

\preprint{}

\title{High-Energy Neutrinos from Cosmic-Ray Scatterings with Supernova Neutrinos}

\author{Gonzalo Herrera}
\affiliation{Center for Neutrino Physics, Department of Physics, Virginia Tech, Blacksburg, VA 24061, USA}
\email{gonzaloherrera@vt.edu}
\author{Shunsaku Horiuchi}
\affiliation{Department of Physics, Institute of Science Tokyo, 2-12-1 Ookayama, Meguro-ku, Tokyo 152-8551, Japan}
\affiliation{Center for Neutrino Physics, Department of Physics, Virginia Tech, Blacksburg, VA 24061, USA}
\affiliation{Kavli IPMU (WPI), UTIAS, The University of Tokyo, Kashiwa, Chiba 277-8583, Japan}
\email{horiuchi@phys.sci.isct.ac.jp}

\begin{abstract}
Cosmic rays scattering with neutrinos produced in supernovae induce a flux of supernova neutrinos boosted to high energies. We calculate the neutrino flux arising from this new mechanism in environments with large cosmic-ray and supernova densities, such as some Active Galactic Nuclei. Under plausible astrophysical conditions, this flux may be detectable with high-energy neutrino telescopes, just considering the proton-neutrino scattering cross section expected in the Standard Model. Furthermore, the center of mass energy of such scatterings can reach $ \sqrt{s} \sim 10-100$ TeV, where the proton-neutrino cross section may be enhanced by new physics such as extra-dimensional theories. The boosted neutrino signal benefits from such an enhancement in the cross section not only at the detection point on Earth, but also at production in astrophysical sources, which allows us to set novel constraints on the ultra-high energy proton-neutrino cross section with neutrino telescopes.
\end{abstract}

\maketitle


\emph{\textbf{Introduction.}}\label{sec:introduction}
It is commonly believed that the production of high-energy astrophysical neutrinos arises predominantly from cosmic-ray proton-proton ($pp$) and cosmic-ray proton-photon ($p\gamma$) collisions in environments where cosmic rays can be accelerated, such as Active Galactic Nuclei \cite{Stecker:1978ah, Stecker:1991vm, Murase:2018utn}\footnote{Note there are alternatives, $\textit{e.g.,}$ an interesting mechanism was proposed in \cite{Hooper:2023ssc}, where very high-energy gamma rays scattering off ambient photons may exceed the threshold for muon production, yielding neutrinos from muon decay. It was recently discussed that this mechanism would however exceed X-ray measurements when applied to NGC 1068 \cite{Das:2024vug}.}. However, recent high-energy neutrino measurements by IceCube and KM3NeT are difficult to accommodate with single-zone leptohadronic models relying on these mechanisms, and standard astrophysical models able to reproduce the observed high-energy neutrino fluxes typically also produce a large flux of X rays and gamma rays sometimes in tension with observations \cite{IceCube:2018dnn, IceCube:2018cha, KM3NeT:2025ccp, Eichmann:2022lxh, Padovani:2018acg,Rodrigues:2018tku,Gao:2018mnu, Murase:2022dog, Neronov:2025jfj,Blanco:2023dfp, Eichmann:2022lxh, Das:2024vug, Fang:2025nzg, Fiorillo:2023dts, Fiorillo:2024akm, Yuan:2025ctq}.

In the same environments where cosmic rays are likely to be accelerated, lower-energy neutrinos with typical energies of $\sim 10$ MeV can be copiously produced in core-collapse supernovae  \cite{Janka:2017vlw, Bahcall:2000nu}. Direct evidence for these MeV neutrinos from core-collapse supernovae was obtained from SN1987A in the Large Magellanic Cloud \cite{Kamiokande-II:1987idp,Bionta:1987qt,Hirata:1988ad,IMB:1988suc}. Recently, Super Kamiokande has reported the first hint of the long sought and predicted Diffuse Supernova Neutrino Background from all redshifts \cite{harada_2024_12726429}, which would directly confirm the ubiquitous production of supernova neutrinos throughout the Universe \cite{Horiuchi:2008jz,Beacom:2010kk,Lunardini:2010ab}.

Here, we propose a new alternative mechanism to produce high-energy neutrinos, namely cosmic rays scattering off supernova neutrinos and boosting supernova neutrinos to higher energies. The analogous process in which cosmic rays boost the cosmic neutrino background to large energies has been previously considered \cite{Ciscar-Monsalvatje:2024tvm,DeMarchi:2024zer, Herrera:2024upj,Zhang:2025rqh}, shown to yield a sizable (ultra)-high energy neutrino flux.  Scatterings of supernova neutrinos would be particularly relevant in environments where the number density of supernova neutrinos is not much smaller than those of ambient protons and photons, or in sources where the typical energy of ambient photons is significantly lower than that of supernova neutrinos. Furthermore, even if the target neutrino density is smaller than those of protons and photons, the boosted neutrino flux may be enhanced at the production point by Beyond the Standard Model deviations from the neutrino-proton cross section. In fact, for a supernova neutrino with energy of $E_{\rm SN\nu} \sim 10$ MeV, the center of mass energy for a collision with a Greisen-Zatsepin-Kuzmin (GZK \cite{Greisen, Zatsepin:1966jv}) proton with energy $T_{p}^{\rm GZK} \simeq 5 \times 10^{10}$ GeV is, 
\begin{equation}
\sqrt{s} \sim \sqrt{2E_{\rm SN\nu}E_{p}} \sim 31.6 \, \mathrm{TeV},
\end{equation}
i.e., energies where the neutrino-proton cross section has not been experimentally measured. Here, the quoted maximum cosmic ray energy in the GZK limit only applies to cosmic-ray protons propagating over extragalactic distances. This number can be larger for heavy nuclei, or cosmic rays produced in certain astrophysical environments, thus possibly reaching even $\sqrt{s} \sim 100$ TeV.

We argue that cosmic ray-neutrino scattering is another independent mechanism to produce high-energy neutrinos that is complementary to the widely explored  $pp$ and $p\gamma$ processes and their final products, as well as the boosted cosmic neutrino background. While the average density of supernova neutrinos is expected to be significantly smaller than the density of cosmic relic neutrinos, we demonstrate here that the suppression in number density can be compensated by the larger center of mass energy of the cosmic ray-supernova neutrino scattering, which enhances the cross section in the Standard Model, in addition to it being possibly further enhanced in Beyond the Standard Model scenarios.

An important aspect for this mechanism is the density of supernova neutrinos. We show that this number can be reasonably large in some active galaxies. The averaged supernova neutrino density over the volume of a galaxy can be estimated from the rate of core-collapse supernovae in that galaxy, $R_{\rm SN\nu} \sim 0.1-10$ yr$^{-1}$ \cite{Van_Den_Bergh_1996}, the number of neutrinos emitted per supernova $N_{\nu} \sim 10^{58}$ \cite{Janka:2017vlw}, and the radial extension of galaxies $r_{\rm gal} \sim 10-100$ kpc as\footnote{Supernova neutrinos typically take a much longer time to escape its host galaxy than it takes for a new supernova to occur: since $\tau_{\rm esc} > r_{\rm gal}/c \simeq 3 \times 10^{5}$yr and $\tau_{\rm SN}= 1/R_{\rm SN} \simeq 1-100$ yr, thus $\tau_{\rm esc} \gg \tau_{\rm SN}$.}
\begin{equation}
\langle n_{\rm SN\nu} \rangle=\frac{R_{\mathrm{SN}} \, N_\nu \,  r_{\mathrm{gal}}}{V_{\mathrm{gal}}\,  c} \sim 10^{-6}-10^{-4} \, \mathrm{cm}^{-3}. 
\end{equation}

While this density is significantly smaller than the average density of the cosmic neutrino background, $n_{\rm C\nu B} \sim 336$ cm$^{-3}$ \cite{Dolgov:1997mb}, the energy of supernova neutrinos is $\sim 10^7$ eV, in contrast to the rest mass of neutrinos from the cosmic neutrino background, smaller than $\sim 0.5$ eV \cite{Katrin:2024tvg}. This crucially affects the production of the boosted neutrino fluxes, since the scattering cross section in the Standard Model in the regime of interest scales with the center of mass energy ($s$). Furthermore, in some Beyond the Standard Model scenarios such as extra-dimensional theories, the cross section scales differently than in the Standard Model \cite{Jain:2000pu, Lykken:2007kp}, which can already be strongly constrained by upper limits on high-energy neutrinos  by IceCube and ANITA.\\

\begin{figure*}[t!]
		\centering
        \includegraphics[width=0.48\textwidth]{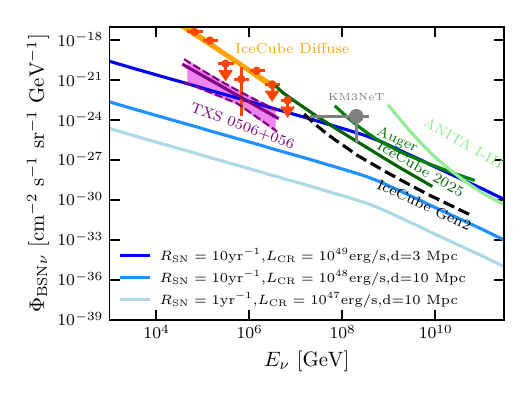}
        \includegraphics[width=0.48\textwidth]{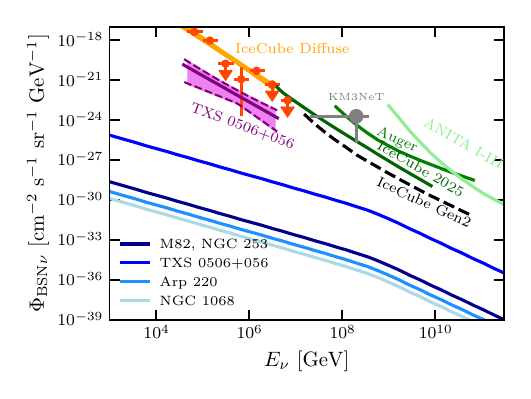}
		\centering
		\caption{\textit{Left panel:} High-energy neutrino flux arising from cosmic-ray scatterings with supernova neutrinos in distant blazars. Adopted values of the supernova rate, cosmic-ray luminosity and distance to the source are labeled; these are chosen from individually plausible value ranges. Shown for comparison are current measurements of diffuse astrophysical neutrinos in IceCube \cite{IceCube:2020acn}, from the galaxy TXS 0506+056 \cite{IceCube:2018cha}, the recent measurement of an ultra-high energy neutrino at KM3Net \cite{KM3NeT:2025ccp, KM3NeT:2025npi}, and upper limits from Auger \cite{PierreAuger:2015ihf}, IceCube \cite{IceCube:2025ezc}, ANITA \cite{ANITA:2018vwl}, and projected limits from IceCube Gen-2 \cite{IceCube-Gen2:2021rkf}. Similar projected bounds to IceCube Gen2 have been derived from GRAND \cite{GRAND:2018iaj}, POEMMA \cite{POEMMA:2020ykm}, and TRIDENT \cite{TRIDENT:2022hql}. \textit{Right panel:} Same as the left plot, except this time considering specific active galaxies and blazars where the necessary astrophysical parameters have been inferred and/or constrained (See Table \ref{tab:cr_sources} for details). For these sources, the expected boosted supernova neutrino flux via Standard Model neutral current scatterings lies at least $\sim 4$ orders of magnitude away from IceCube upper limits.}
  \label{fig:boostedSN}
\end{figure*}

\emph{\textbf{Boosted supernova neutrino flux.}}\label{sec:fluxes}
We compute the cosmic-ray boosted supernova neutrino flux from a distant galaxy reaching the Earth as,
\begin{align}
\frac{d \Phi_{\mathrm{BSN}\nu}}{d E_\nu}=& \frac{1}{4 \pi d^2}\int_{R_{\rm acc}}^{r_{\rm gal}} dr \, 4\pi r^2 \nonumber \\ & \times\int_{T_{\mathrm{p}}^{\min }\left(E_\nu\right)}^{T_{\mathrm{p}}^{\mathrm{max}}} d T_{\mathrm{p}} n_{\rm SN\nu}(r)\, \frac{d \Phi_{\mathrm{p}}}{d T_{\mathrm{p}}d\Omega}(r) \frac{d \sigma_{\nu \mathrm{p}}}{d E_\nu},
\end{align}
where $d$ is the distance from the source to Earth, $T_{p}^{\rm min} (E_{\nu})$ is the minimum cosmic-ray energy needed to induce a given neutrino energy $E_\nu$ as allowed by the kinematics of the scattering. $T_{p}^{\rm max}$ is fixed by the maximum cosmic ray energy, which we take to be $T_{p}^{\rm max}=10\,T_{p}^{\rm GZK}=5 \times 10^{11}$ GeV. $R_{\rm acc}$ is the cosmic-ray acceleration radius, typically much smaller than the size of the galaxy $r_{\rm gal}$; we adopt as fiducial $R_{\rm acc}=0.1$ pc \cite{Padovani:2019xcv}. We checked that smaller values as those inferred for some sources like NGC 1068 do not modify the boosted fluxes appreciably for our assumed cosmic-ray and supernova radial profiles \cite{Murase:2022dog}. 

Below, we consider cosmic rays in two kinds of active galaxies: jetted (blazar-like) and non-jetted. For the jetted, Eq.~(\ref{eq:blazar_flux}), the volume factor $4\pi r^2$ cancels the $1/r^2$ in $n_{\rm SN\nu}(r)$ analytically and the flux is insensitive to $R_{\rm acc}$ at the $R_{\rm acc}/r_{\rm gal}\sim 10^{-5}$ level. For the non-jetted, Eq.~(\ref{eq:CR_spectra}), the integrand scales as $r^{-3/2}$ and the flux therefore depends on the inner cutoff approximately as $\Phi_{\rm BSN\nu}\propto R_{\rm acc}^{-1/2}$; smaller values of $R_{\rm acc}$ enhance the flux, larger values reduce it. For plausible values spanning $R_{\rm acc}\sim 10^{-3}$--$10$~pc (bracketing AGN coronae, blazar zone scales, and more extended starburst acceleration regions), this translates into a factor $\sim 10$ enhancement at $R_{\rm acc}\sim 10^{-3}$~pc and a factor $\sim 10$ suppression at $R_{\rm acc}\sim 10$~pc relative to our fiducial $R_{\rm acc}=0.1$~pc, i.e., a total factor $\sim 100$ across the astrophysically motivated range. This $R_{\rm acc}$ systematic is of comparable order to the astrophysical uncertainties on $L_{\rm CR}$ and $R_{\rm SN}$ discussed above, and should be combined with them when folding all astrophysical inputs into the predicted flux. 

The cosmic-ray flux at the source, $d \Phi_{\mathrm{p}}/d T_{\mathrm{p}}$, is assumed to be purely composed of protons. Although at the highest energies above $\sim 10^{10}$ GeV there is increasing evidence for the composition to depart from pure protons to nuclei, $\textit{e.g.,}$ \cite{AlvesBatista:2019tlv,Abreu:2025ivn}, we will find that the most detectable contribution to the boosted supernova neutrinos arises mostly from just below the transition, i.e., $\sim 10^9$--$10^{10}$ GeV, where recent upper limits from IceCube are most sensitive \cite{IceCube:2025ezc}. We model the cosmic-ray flux differently for each type of source considered. For active galaxies with no evidence for a jet ($\textit{e.g.,}$ M82, NGC 253, Arp 220, NGC 1068), we model the cosmic-ray flux isotropically with a power-law in energy given by,
\begin{equation}
\frac{d \Phi_{\rm p}(T_p,r)}{d T_{p}d\Omega}=
		   \frac{L_p}{4 \pi r^2 m_p^2} \, \left(\frac{T_p}{m_p}\right)^{-\alpha},
	\label{eq:CR_spectra}
\end{equation}
where $T_p=E_p-m_p$ is the kinetic energy of the cosmic-ray protons, $L_p$ is the cosmic-ray proton luminosity at the source considered, $m_p$ is the proton mass and $\alpha$ is the spectral index. We adopt a modestly concentrated cosmic-ray distribution towards the center of the galaxy following $\propto r^{1/2}$, which is motivated by the distributions of cosmic-ray sources being radially concentrated, $\textit{e.g.,}$ see detailed propagation studies in the Milky Way galaxy \cite{Porter:2017vaa,Johannesson:2018bit}. For our benchmark calculations, we take $\alpha=2$ for all sources considered in this work. 

For blazars ($\textit{e.g.,}$ TXS 0506+056), we model the cosmic-ray flux following \cite{Gorchtein:2010xa,Wang:2021jic, Granelli:2022ysi} as,
\begin{align}
\frac{d \Phi_p (T_p,r)}{d T_p d \Omega}= & \frac{1}{4 \pi} c_p\left(1+\frac{T_p}{m_p}\right)^{-\alpha} \nonumber \\
& \times \frac{\beta_p\left(1-\beta_p \beta_B \mu\right)^{-\alpha} \Gamma_B^{-\alpha}}{\sqrt{\left(1-\beta_p \beta_B \mu\right)^2-\left(1-\beta_p^2\right)\left(1-\beta_B^2\right)}},
\label{eq:blazar_flux}
\end{align}
where the proton velocity is 
\begin{equation}
\beta_p=\left[1-m_p^2 /\left(T_p+m_p\right)^2\right]^{1/2}.
\end{equation}
Furthermore, $\gamma_{\min , p}^{\prime}$ and $\gamma_{\max , p}^{\prime}$ denote the minimal and maximal Lorentz factors,  $\mathcal{D}=\left[\Gamma_B\left(1-\beta_B \cos \theta_{\text {l.o.s.}}\right)\right]^{-1}$ denotes the Doppler factor, and $c_p$ can be obtained from $L_p$ \cite{Gorchtein:2010xa,Wang:2021jic}. Here we only consider one blazar, TXS 0506+056, and take the parameters from \cite{Cerruti:2018tmc}.

The $d \sigma_{\nu \mathrm{p}}/d E_\nu$ is the differential proton-neutrino cross section in the neutrino rest frame. In the Standard Model, the neutral current elastic cross section (averaged over neutrinos and antineutrinos) reads \cite{Giunti:2007ry},
\begin{equation}
\frac{d \sigma_{\nu p}^{\mathrm{elastic}}}{d E_\nu}=\frac{2 G_F^2 m_\nu m_p^4}{\pi\left(s-m_p^2\right)^2}\left[A_p\left(Q^2\right)+C_n\left(Q^2\right) \frac{(s-u)^2}{m_p^4}\right],
\end{equation}
where we take the neutral current form factors $A$ and $C$ from \cite{Giunti:2007ry, Formaggio:2012cpf}. Beyond center of mass energies of $\sqrt{s} \gtrsim 1$ GeV, Deep Inelastic Scattering (DIS) takes over the elastic cross section. In this regime, we extract the scattering cross section and inelasticity of the scattering process from NUANCE \cite{Casper:2002sd,Formaggio:2012cpf}, averaging over neutrinos and antineutrinos. Beyond center of mass energies of $\sqrt{s} \gtrsim 4$ TeV, the $Z$ boson propagator suppresses the cross section, so we match the DIS cross section with a cross section scaling with center of mass energy of $s^{0.363}$, as inferred in \cite{Gandhi:1995tf}. This description adequately captures the energy dependence and normalization of the neutral current cross section in the Standard Model in all energy regimes of interest.

Finally, we are left to describe the distribution of supernova neutrinos. We use
\begin{equation}
n_{\rm SN\nu}(r)=f_{\rm SN}(r)\frac{R_{\mathrm{SN}} \, N_\nu}{4 \pi r^2 \,c},
\end{equation}
where all variables have been previously introduced, except the radial distribution of supernovae $f_{\rm SN}$. We use an exponential distribution peaked at the center of the Galaxy under consideration as \cite{Verberne:2021tse, Ranasinghe:2022ntj}
\begin{equation}\label{eq:SN_distribution}
f_{\rm SN}(r)=\exp \left(-\frac{r}{r_{\rm gal}}\right).
\end{equation}

\begin{table}[t!]
\centering
\begin{tabular}{|l|c|c|c|c|}
\hline
Galaxy & d (Mpc) & $R_{\rm SN} (\mathrm{yr}^{-1})$ & $L_{\rm CR}$ (erg/s) \\
\hline
M82 \cite{Yoast-Hull:2013wwa}     & 3.3   & $\lesssim 1$          &  $\lesssim 10^{45}$ \\     
NGC 253 \cite{1984MNRAS.207..671B, Rampadarath:2013vna}            & 3.5   & $\lesssim 1$          & $\lesssim10^{45}$   \\
NGC 1068 \cite{Murase:2022dog}            & 14   & $\lesssim 1$          & $\lesssim 10^{45}$   \\
Arp 220 \cite{Peng:2016nsx}            & 77    & $\lesssim 10$         & $ \lesssim 10^{46}$      \\
TXS 0506+056 \cite{Padovani:2019xcv}            & 1762    & $\lesssim 10$         & $ \lesssim 10^{49}$      \\
\hline
\end{tabular}
\caption{Upper limits on some astrophysical parameters for the sources considered in this in this work.}
\label{tab:cr_sources}
\end{table}

\vspace{3mm}
\begin{figure*}[t!]
		\centering
        \includegraphics[width=0.48\textwidth]{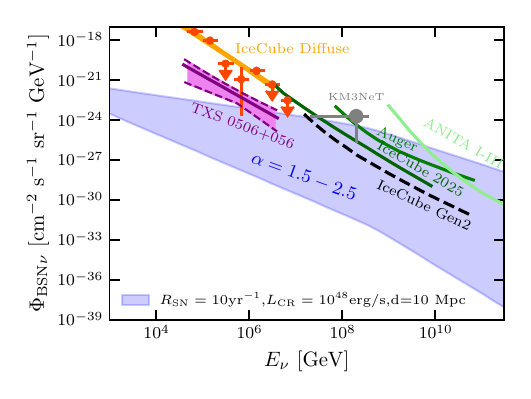}
        \includegraphics[width=0.48\textwidth]{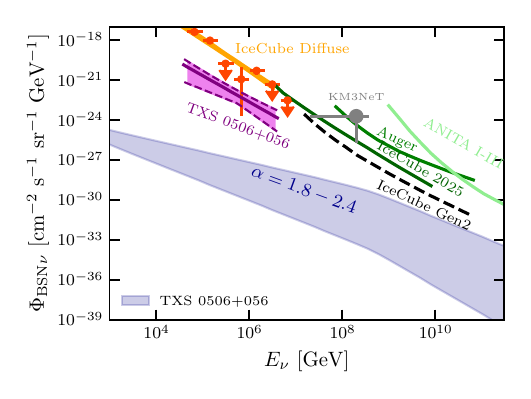}
		\centering
		\caption{\textit{Left panel:} Dependence of the cosmic-ray spectral index ($\alpha$) on the boosted supernova neutrino flux, using benchmark astrophysical parameters as labeled. \textit{Right panel:} Dependence of the cosmic-ray spectral index on the boosted supernova neutrino flux from TXS 0506+056, where the spectral index range is taken from the time-dependent analysis of IceCube data \cite{IceCube:2018cha}.}
  \label{fig:alpha}
\end{figure*}

Combining the above prescriptions, we can compute the high energy neutrino flux arising from cosmic-ray scatterings with supernova neutrinos via neutral current interactions in distant galaxies. The main astrophysical uncertain parameters in our calculation are the cosmic-ray luminosity $L_p$, the spectral index $\alpha$, and the supernova rate $R_{\rm SN}$. While the distance to the source $d$ is also important, typically distances to galaxies are better established. 

In the left panel of Fig.~\ref{fig:boostedSN}, we show some optimistic yet plausible examples for values of these parameters that can be realized in Nature. For instance, cosmic-ray luminosities as large as $L_{\rm CR} \sim 10^{49}$ erg/s have been inferred from TXS 0506+056 \cite{IceCube:2018cha, IceCube:2018dnn, Rodrigues:2018tku}, supernova rates as large as $R_{\rm SN} \sim 10$ yr$^{-1}$ have been inferred from Arp 220 \cite{Peng:2016nsx}, and powerful starburst galaxies reside at distances as close as $d \simeq 3$ Mpc \cite{Yoast-Hull:2013wwa,1984MNRAS.207..671B}.
For this plot, we consider the cosmic-ray flux model for blazars, described by Eq.~(\ref{eq:blazar_flux}). Our simulations demonstrate that cosmic rays can boost supernova neutrinos, yielding a potentially detectable flux on Earth. We emphasize, however, that no single known blazar simultaneously realizes all the parameters chosen in this illustrative envelope, so this panel should be interpreted as a parameter study rather than a prediction for a specific known source. The boosted flux scales as $\Phi_{\rm BSN\nu}\propto L_{\rm CR}\,R_{\rm SN}/d^2$, allowing rescaling to any specific source.

Predictions for a sample of currently known sources for which the maximum supernova rate and cosmic-ray luminosities are available are shown in the right panel of Fig.~\ref{fig:boostedSN} and in Table \ref{tab:cr_sources}. The figure shows the boosted supernova neutrino fluxes for the known parameters, as labeled. The fluxes are below current experimental sensitivities by about 4 to 8 orders of magnitude, depending on the source considered. The most sensitive energy range are those where IceCube provides the best searches \cite{IceCube:2025ezc}. The cosmic-ray energies responsible for these boosts are $\sim 10^8$--$10^{10}$ GeV, justifying our assumption of proton composition. We find that blazars like TXS 0506+056 are expected to provide the largest boosted supernova neutrino fluxes, due to the enhancement in the cosmic-ray flux due to jet collimation along the line of sight. We note that while small, the boosted supernova neutrino flux is comparable to or larger than the cosmic-ray boosted cosmic neutrino background in similar sources \cite{Herrera:2024upj, Zhang:2025rqh}. 

While the up-scattered neutrino flux may be below experimental sensitives in the above examples, they can nonetheless be key in environments where, in the region where cosmic rays are accelerated, the supernova-neutrino density is comparable to or larger than the gas and photon densities driving conventional $pp$ and $p\gamma$ neutrino production. Relevant examples include young compact massive-star clusters accelerating cosmic rays at collective wind termination shocks inside super-bubbles swept clean by repeated supernovae~\cite{Bykov:2014lpa,Aharonian:2018oau}; the termination shocks of starburst-driven galactic superwinds, far from the dense central star-forming region but still embedded in the supernova-neutrino reservoir of the host galaxy~\cite{Strickland:2009ft,HeckmanThompson2017}; young supernova remnants expanding into pre-existing low-density cavities, in which the very supernova explosion responsible for cosmic-ray acceleration also supplies a high local supernova-neutrino density; and Population~III and early-Universe supernova environments, where the pre-enrichment interstellar medium carries a much smaller baryon and photon opacity than its low-redshift counterparts. In such systems the relative weight of the boosted channel is enhanced and warrants dedicated source-specific modelling of the co-spatial cosmic-ray, gas, photon, and supernova-neutrino distributions.

We explore the impact of the cosmic-ray spectral index on the ensuing high-energy neutrino fluxes from cosmic-ray scatterings with supernova neutrinos. We take a possible range of values $\alpha=1.5$--$2.5$, and illustrate the impact of this choice fixing the supernova rate to $R_{\rm SN} = 10$ yr$^{-1}$, cosmic-ray luminosity to $L_{\rm CR} = 10^{48}$, and distance to $d=10$ Mpc; the result is shown by a blue band in the left panel of Fig.~\ref{fig:alpha}. Under the blazar cosmic-ray model from Eq.~(\ref{eq:blazar_flux}), the spectral index can affect the high-energy neutrino fluxes by orders of magnitude, especially at the highest energies. In the right panel of the Figure, we illustrate the impact of the cosmic-ray spectral index for the predicted fluxes from TXS 0506+056, where the uncertain range ($\alpha=1.8-2.4$) is taken from the time-dependent analysis of IceCube data in \cite{IceCube:2018cha}.
\\

\emph{\textbf{Upper limits on the ultra-high energy proton-neutrino cross section.}}\label{sec:BSM}

In some Beyond the Standard Model scenarios, the proton-neutrino cross section could scale with energy differently at high energies than $\sigma \sim s^{0.363}$ or $\sigma \sim s$, as expected in the Standard Model. One yet largely unconstrained possibility arises from extra-dimensional theories, where the cross section can scale as $\sigma \sim s^2$ for energies larger than a fundamental scale $M_{\star}$ \cite{Jain:2000pu,Lykken:2007kp, Esteban:2022uuw}. This scale can mean different things depending on the extra-dimensional theory under consideration. Typically, they indicate the energy scale at which graviton exchange becomes strong, such as in Kaluza-Klein models. If such scale were as low as a few hundreds of TeV, it would enhance the cosmic-ray boosted supernova neutrino fluxes both at the source production, and the detection on Earth-based high-energy neutrino experiments. We study this possibility and derive limits on the scale $M_{\star}$ from the non-observation of a supernova boosted high-energy neutrino flux from TXS 0506+056. Weaker limits would be obtained for other Galaxies, see Table \ref{tab:cr_sources} for a few representative examples.

\begin{figure}[t!]
		\centering
        \includegraphics[width=0.48\textwidth]{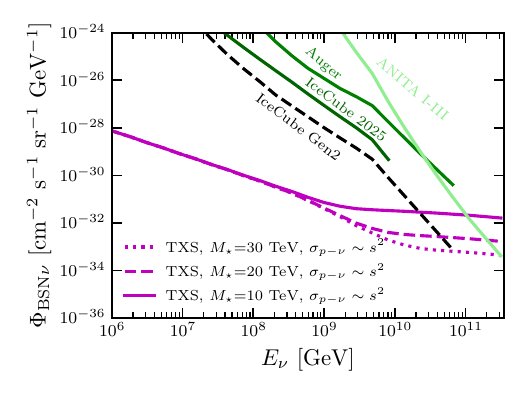}
		\centering
		\caption{High energy neutrino flux from cosmic-ray scatterings with supernova neutrinos in TXS 0506+056, for different values of a new physics scale $M_{\star}$, where the cross section begins to scale as $\sigma_{p-\nu} \sim s^2$ (see main text for details). The upper limits from high-energy neutrino experiments are rescaled to account for the enhanced cross section at high center of mass energies. We note that a scale $M_{\star} \lesssim 30$ TeV would lead to a boosted supernova neutrino flux overcoming limits from current experiments. The strongest limit arises from ANITA \cite{ANITA:2018vwl}.}
  \label{fig:BSM}
\end{figure}

We illustrate the enhancement on the boosted supernova neutrino fluxes in Fig.~\ref{fig:BSM} for various new physics energy scale $M^{\star}$, compared to measurements and limits from high-energy neutrino experiments. Due to the enhancement of the cross section at high energies, we recast the upper limits from experiments at each center of mass energy by a factor $\sigma^{\rm SM}(s)/\sigma^{\rm BSM}(s^2)$, which strengthens the upper limits on the high-energy neutrino flux. It can be seen in the plot that values of $M_{\star} \lesssim 30$ TeV would lead to a cosmic-ray boosted supernova neutrino flux that overshoots the upper limits from ANITA. 

Our results place a new strong bound on the scale of extra-dimensional theories $M_{\star}$, the first one arising from high-energy neutrino experiments, and comparable with complementary bounds. For instance, the previously most stringent astrophysical bound arises from graviton production in SN1987A, which restricts $M_{\star}>27$ TeV \cite{Hanhart:2001fx}, comparable to our bound. Cosmological bounds from graviton decay contributions to the diffuse gamma-ray background have also been derived, being as stringent as $M_{\star} > 100$ TeV. It has been discussed that these bounds are relaxed if gravitons can also decay on other branes \cite{Hall:1999mk}. Collider bounds on extra-dimensions can also be strong, as stringent as $M_{\star} \gtrsim 10$ TeV \cite{CMS:2017zts}.\\
\\ 

The bound we quote assumes $L_{\rm CR}\sim 10^{49}$~erg/s for TXS~0506+056, a value consistent with the range $L_{\rm CR}\sim 10^{48}$--$10^{49}$~erg/s inferred from published multi-wavelength and multi-messenger fits of the 2014--2015 and 2017 neutrino observations~\cite{Cerruti:2018tmc,Rodrigues:2018tku,IceCube:2018cha,IceCube:2018dnn}. In the BSM regime $\sigma_{p\nu}\propto s^2/M_\star^4$, so the predicted boosted flux scales as $L_{\rm CR}/M_\star^4$ and the corresponding bound scales as $M_\star\propto L_{\rm CR}^{1/4}$. An order-of-magnitude uncertainty in $L_{\rm CR}$ therefore translates into a factor $\simeq 1.8$ uncertainty in $M_\star$, which is smaller than the intrinsic modelling uncertainty of the complementary bounds listed above. This dependence propagates linearly through all subsequent quotes of the $M_\star$ bound.

\emph{\label{sec:conclusions}
\textbf{Conclusions.}}
The origin of (ultra)high-energy neutrinos is widely discussed to be linked to cosmic-ray acceleration and subsequent $p\gamma$ and $pp$ interactions. However, the combined data from high-energy neutrino experiments and gamma-ray telescopes disfavors the simplest single-zone leptohadronic models. We have proposed an alternative mechanism to produce (ultra)high-energy neutrinos via Standard Model processes, arising from the scatterings of cosmic rays with supernova neutrinos (see Fig.~\ref{fig:boostedSN}). This effect becomes efficient at producing a detectable neutrino flux for large cosmic-ray luminosities (but still below the Eddington limit of known sources), nearby distances (but within distances where powerful galaxies have been observed), and high supernova rates (but within rates of nearby starbursting galaxies). However, when combining these ingredients for a sample of existing sources with parameter estimates (e.g., M82, NGC 1068, Arp 220 and TXS 0506+056; see Table \ref{tab:cr_sources}), we predict fluxes that lie between $\sim 4-8$ orders of magnitude below current sensitivities of high-energy neutrino experiments, only aligning with some pessimistic cosmogenic neutrino models \cite{Kotera_2010}. We note, however, the strong dependence on the cosmic-ray energy spectrum, which can strongly enhance the flux for harder spectral indices (see Fig.~\ref{fig:alpha}). 

Our prediction could be larger for additional reasons. For example, the above conclusion is true only if the intrinsic ultra high energy cosmic-ray luminosity at these sources is smaller than the observed bolometric luminosity. The intrinsic cosmic-ray luminosity could be larger, but be hindered by a large proton or photon opacity. Additionally, we have assumed a radially concentrated cosmic-ray density and supernova rate profiles, which gives an enhancement in the boosted supernova fluxes of $\sim 16.7$ compared to simply taking the product of the average values of both quantities within the Galaxy. However, the cosmic-ray profile in particular remains uncertain, in particular in light of unknown propagation parameters, which may end up yielding a cosmic-ray spatial profile differing from the prescription we followed. This task is left for future investigation. Finally, we have not included the effects of neutrinos emitted in core collapse that do not yield supernovae, i.e., collapse to black holes with supernovae. In these systems, the emitted neutrinos are systematically of higher energies ($\textit{e.g.,}$ \cite{Liebendoerfer:2002xn}), therefore are more likely to be boosted due to the higher cross section. While the fraction of stars undergoing the black hole channel remains uncertain, it is plausibly in the few tens of percent \cite{Horiuchi:2014ska,Neustadt:2021jjt}, making them potentially non-negligible.

In addition to the galaxies shown in Table \ref{tab:cr_sources}, all galaxies hosting supernovae and cosmic rays should yield a boosted supernova neutrino flux. However, the galaxies shown in Table \ref{tab:cr_sources} are not necessarily representative and the calculation of the cosmic-ray boosted supernova neutrino background is left for future work. In addition, one may also consider a diffuse population of stellar neutrinos which cosmic rays scatter off. The number density of stellar neutrinos in galaxies can be comparable to that of supernova neutrinos, but their energy is lower, so we expect this contribution to be sub-dominant compared to the cosmic-ray boosted supernova neutrinos. A complementary extension is to consider a source-correlated, higher-energy neutrino target; for example the conventional $pp$ or $p\gamma$ neutrino field co-spatial with the cosmic-ray accelerator. In such a setting the target number density scales as $n_\nu\simeq L_\nu/(4\pi R^2 c\,\bar{E}_\nu)$, and the source-side interaction fraction $f_{p\nu}\simeq L_\nu\,\sigma_{p\nu}/(4\pi R c\,\bar{E}_\nu)$ requires a cross section $\sigma_{p\nu}^{\rm req}\simeq 4\pi R c\,\bar{E}_\nu f_{p\nu}/L_\nu$. Moving the target from $\bar{E}_\nu\sim 10$~MeV (our supernova-neutrino case) to $\bar{E}_\nu\sim 10$~TeV--$1$~PeV at fixed $L_\nu$ reduces $n_\nu$ by $\sim 10^6$--$10^8$, but raises the center-of-mass energy of the scattering from $\sqrt{s}\sim 30$~TeV to $\sqrt{s}\sim 100$--$500$~PeV, a regime inaccessible to direct $\nu N$ measurements. For compact AGN coronae this yields $\sigma_{p\nu}^{\rm req}\sim 10^{-19}$--$10^{-18}$~cm$^2$ (for $f_{p\nu}\sim 10^{-2}$), while for compact, neutrino-luminous environments such as choked jets or failed low-luminosity gamma-ray bursts~\cite{Meszaros:2001ms,Murase:2013ffa,Senno:2015tsn} the required cross section falls into the hadronic-like window $\sigma_{p\nu}^{\rm req}\sim 10^{-26}$--$10^{-24}$~cm$^2$, still far above the Standard Model but corresponding to an unexplored $s$ regime where new strong dynamics could plausibly turn on. A non-observation of the boosted signal from such sources would therefore translate into the first astrophysical upper bound on $\sigma_{p\nu}$ at $\sqrt{s}\gtrsim 10$~PeV, complementary to the extra-dimensional bound derived here from the diffuse supernova-neutrino target. Translated into a constraint on the extra-dimensional scale via the $\sigma_{p\nu}\propto s^2/M_\star^4$ enhancement, non-observation of the boosted signal from such compact, neutrino-luminous sources would yield $M_\star\gtrsim 10$~TeV for AGN-corona benchmarks (weaker than our diffuse-target bound) up to $M_\star\gtrsim 100$~TeV for choked-jet or failed low-luminosity gamma-ray burst benchmarks (one to two orders of magnitude stronger than the diffuse-target bound), depending on the source parameters. A detailed source-specific treatment is left for future work.

We have further pointed out that the scattering of cosmic rays with supernova neutrinos can reach center of mass energies as large as $\sqrt{s} \sim 100$ TeV, where the neutrino-proton cross section has not been measured and enhancements induced by Beyond the Standard Model effects are possible. Interestingly, this enhancement would not only manifest at detection on Earth-based experiments, but also at the production point in extragalactic sources. In fact, for the blazar TXS 0506+056, we find that the non-observation of ultra-high energy neutrinos with ANITA allows to place an upper limit on the scale of extra-dimensions of $M_{\star} \gtrsim 30$ TeV (see Fig.~\ref{fig:BSM}), comparable to complementary probes, and the first one obtained with astrophysical high-energy neutrino data.

\section*{Acknowledgments}
We thank Bhupal Dev, Matheus Hostert and Stephan Meigen Berger for useful discussions. G.H.~is supported by the U.S. Department of Energy under the award number DE-SC0020262. The work of S.H.~is supported by NSF Grant No.~PHY-2209420 and JSPS KAKENHI Grant Number JP22K03630 and JP23H04899. This work was supported by World Premier International Research Center Initiative (WPI Initiative), MEXT, Japan. 
\noindent

\bibliography{references}

@article{Ciscar-Monsalvatje:2024tvm,
    author = "C\'\i{}scar-Monsalvatje, Mar and Herrera, Gonzalo and Shoemaker, Ian M.",
    title = "{Upper limits on the cosmic neutrino background from cosmic rays}",
    eprint = "2402.00985",
    archivePrefix = "arXiv",
    primaryClass = "hep-ph",
    doi = "10.1103/PhysRevD.110.063036",
    journal = "Phys. Rev. D",
    volume = "110",
    number = "6",
    pages = "063036",
    year = "2024"
}

@article{AlvesBatista:2019tlv,
    author = "Alves Batista, Rafael and others",
    title = "{Open Questions in Cosmic-Ray Research at Ultrahigh Energies}",
    eprint = "1903.06714",
    archivePrefix = "arXiv",
    primaryClass = "astro-ph.HE",
    doi = "10.3389/fspas.2019.00023",
    journal = "Front. Astron. Space Sci.",
    volume = "6",
    pages = "23",
    year = "2019"
}

@article{Horiuchi:2014ska,
    author = "Horiuchi, Shunsaku and Nakamura, Ko and Takiwaki, Tomoya and Kotake, Kei and Tanaka, Masaomi",
    title = "{The red supergiant and supernova rate problems: implications for core-collapse supernova physics}",
    eprint = "1409.0006",
    archivePrefix = "arXiv",
    primaryClass = "astro-ph.HE",
    doi = "10.1093/mnrasl/slu146",
    journal = "Mon. Not. Roy. Astron. Soc.",
    volume = "445",
    pages = "L99",
    year = "2014"
}

@article{Liebendoerfer:2002xn,
    author = "Liebendoerfer, M. and Messer, O. E. B. and Mezzacappa, A. and Bruenn, S. W. and Cardall, C. Y. and Thielemann, F. K.",
    title = "{A Finite difference representation of neutrino radiation hydrodynamics for spherically symmetric general relativistic supernova simulations}",
    eprint = "astro-ph/0207036",
    archivePrefix = "arXiv",
    doi = "10.1086/380191",
    journal = "Astrophys. J. Suppl.",
    volume = "150",
    pages = "263--316",
    year = "2004"
}

@article{Neustadt:2021jjt,
    author = "Neustadt, J. M. M. and Kochanek, C. S. and Stanek, K. Z. and Basinger, C. M. and Jayasinghe, T. and Garling, C. T. and Adams, S. M. and Gerke, J.",
    title = "{The search for failed supernovae with the Large Binocular Telescope: a new candidate and the failed SN fraction with 11~yr of data}",
    eprint = "2104.03318",
    archivePrefix = "arXiv",
    primaryClass = "astro-ph.SR",
    doi = "10.1093/mnras/stab2605",
    journal = "Mon. Not. Roy. Astron. Soc.",
    volume = "508",
    number = "1",
    pages = "516--528",
    year = "2021"
}

@inproceedings{Abreu:2025ivn,
    author = "Abreu, Pedro and others",
    title = "{Measurement and Interpretation of UHECR Mass Composition at the Pierre Auger Observatory}",
    booktitle = "{39th International Cosmic Ray Conference}",
    eprint = "2507.10292",
    archivePrefix = "arXiv",
    primaryClass = "astro-ph.HE",
    reportNumber = "PoS (ICRC2025) 331",
    month = "7",
    year = "2025"
}

@article{Johannesson:2018bit,
    author = "J{\'o}hannesson, Gu{\dj}laugur and Porter, Troy A. and Moskalenko, Igor V.",
    title = "{The Three-Dimensional Spatial Distribution of Interstellar Gas in the Milky Way: Implications for Cosmic Rays and High-Energy Gamma-Ray Emissions}",
    eprint = "1802.08646",
    archivePrefix = "arXiv",
    primaryClass = "astro-ph.HE",
    doi = "10.3847/1538-4357/aab26e",
    journal = "Astrophys. J.",
    volume = "856",
    number = "1",
    pages = "45",
    year = "2018"
}

@article{Porter:2017vaa,
    author = "Porter, Troy A. and Johannesson, Gudlaugur and Moskalenko, Igor V.",
    title = "{High-Energy Gamma Rays from the Milky Way: Three-Dimensional Spatial Models for the Cosmic-Ray and Radiation Field Densities in the Interstellar Medium}",
    eprint = "1708.00816",
    archivePrefix = "arXiv",
    primaryClass = "astro-ph.HE",
    doi = "10.3847/1538-4357/aa844d",
    journal = "Astrophys. J.",
    volume = "846",
    number = "1",
    pages = "67",
    year = "2017"
}

@misc{harada_2024_12726429,
  author       = {Harada, Masayuki},
  title        = {Review of diffuse SN neutrino background},
  month        = jun,
  year         = 2024,
  publisher    = {Zenodo},
  doi          = {10.5281/zenodo.12726429},
  url          = {https://doi.org/10.5281/zenodo.12726429},
}

@article{Bionta:1987qt,
    author = "Bionta, R. M. and others",
    title = "{Observation of a Neutrino Burst in Coincidence with Supernova SN 1987a in the Large Magellanic Cloud}",
    reportNumber = "UCI-NEUTRINO-87-10",
    doi = "10.1103/PhysRevLett.58.1494",
    journal = "Phys. Rev. Lett.",
    volume = "58",
    pages = "1494",
    year = "1987"
}

@article{Lunardini:2010ab,
    author = "Lunardini, Cecilia",
    title = "{Diffuse supernova neutrinos at underground laboratories}",
    eprint = "1007.3252",
    archivePrefix = "arXiv",
    primaryClass = "astro-ph.CO",
    doi = "10.1016/j.astropartphys.2016.02.005",
    journal = "Astropart. Phys.",
    volume = "79",
    pages = "49--77",
    year = "2016"
}

@article{IMB:1988suc,
    author = "Bratton, C. B. and others",
    collaboration = "IMB",
    title = "{Angular Distribution of Events From Sn1987a}",
    reportNumber = "UM-PDK-88-1",
    doi = "10.1103/PhysRevD.37.3361",
    journal = "Phys. Rev. D",
    volume = "37",
    pages = "3361",
    year = "1988"
}

@article{Hirata:1988ad,
    author = "Hirata, K. S. and others",
    title = "{Observation in the Kamiokande-II Detector of the Neutrino Burst from Supernova SN 1987a}",
    doi = "10.1103/PhysRevD.38.448",
    journal = "Phys. Rev. D",
    volume = "38",
    pages = "448--458",
    year = "1988"
}

@article{GRAND:2018iaj,
    author = "\'Alvarez-Mu\~niz, Jaime and others",
    collaboration = "GRAND",
    title = "{The Giant Radio Array for Neutrino Detection (GRAND): Science and Design}",
    eprint = "1810.09994",
    archivePrefix = "arXiv",
    primaryClass = "astro-ph.HE",
    doi = "10.1007/s11433-018-9385-7",
    journal = "Sci. China Phys. Mech. Astron.",
    volume = "63",
    number = "1",
    pages = "219501",
    year = "2020"
}

@article{POEMMA:2020ykm,
    author = "Olinto, A. V. and others",
    collaboration = "POEMMA",
    title = "{The POEMMA (Probe of Extreme Multi-Messenger Astrophysics) observatory}",
    eprint = "2012.07945",
    archivePrefix = "arXiv",
    primaryClass = "astro-ph.IM",
    doi = "10.1088/1475-7516/2021/06/007",
    journal = "JCAP",
    volume = "06",
    pages = "007",
    year = "2021"
}

@article{TRIDENT:2022hql,
    author = "Ye, Z. P. and others",
    collaboration = "TRIDENT",
    title = "{A multi-cubic-kilometre neutrino telescope in the western Pacific Ocean}",
    eprint = "2207.04519",
    archivePrefix = "arXiv",
    primaryClass = "astro-ph.HE",
    doi = "10.1038/s41550-023-02087-6",
    journal = "Nature Astron.",
    volume = "7",
    number = "12",
    pages = "1497--1505",
    year = "2023"
}

@article{Janka:2017vlw,
    author = "Janka, H. -Th.",
    title = "{Neutrino Emission from Supernovae}",
    eprint = "1702.08713",
    archivePrefix = "arXiv",
    primaryClass = "astro-ph.HE",
    doi = "10.1007/978-3-319-21846-54",
    month = "2",
    year = "2017"
}

@article{Bahcall:2000nu,
    author = "Bahcall, John N. and Pinsonneault, M. H. and Basu, Sarbani",
    title = "{Solar models: Current epoch and time dependences, neutrinos, and helioseismological properties}",
    eprint = "astro-ph/0010346",
    archivePrefix = "arXiv",
    doi = "10.1086/321493",
    journal = "Astrophys. J.",
    volume = "555",
    pages = "990--1012",
    year = "2001"
}

@article{Kamiokande-II:1987idp,
    author = "Hirata, K. and others",
    editor = "Wali, K. C.",
    collaboration = "Kamiokande-II",
    title = "{Observation of a Neutrino Burst from the Supernova SN 1987a}",
    reportNumber = "UT-ICEPP-87-01, UPR-142E",
    doi = "10.1103/PhysRevLett.58.1490",
    journal = "Phys. Rev. Lett.",
    volume = "58",
    pages = "1490--1493",
    year = "1987"
}

@article{Horiuchi:2008jz,
    author = "Horiuchi, Shunsaku and Beacom, John F. and Dwek, Eli",
    title = "{The Diffuse Supernova Neutrino Background is detectable in Super-Kamiokande}",
    eprint = "0812.3157",
    archivePrefix = "arXiv",
    primaryClass = "astro-ph",
    doi = "10.1103/PhysRevD.79.083013",
    journal = "Phys. Rev. D",
    volume = "79",
    pages = "083013",
    year = "2009"
}

@article{Beacom:2010kk,
    author = "Beacom, John F.",
    title = "{The Diffuse Supernova Neutrino Background}",
    eprint = "1004.3311",
    archivePrefix = "arXiv",
    primaryClass = "astro-ph.HE",
    doi = "10.1146/annurev.nucl.010909.083331",
    journal = "Ann. Rev. Nucl. Part. Sci.",
    volume = "60",
    pages = "439--462",
    year = "2010"
}

@article{Herrera:2024upj,
    author = "Herrera, Gonzalo and Horiuchi, Shunsaku and Qi, Xiaolin",
    title = "{Diffuse boosted cosmic neutrino background}",
    eprint = "2405.14946",
    archivePrefix = "arXiv",
    primaryClass = "hep-ph",
    doi = "10.1103/PhysRevD.111.063016",
    journal = "Phys. Rev. D",
    volume = "111",
    number = "6",
    pages = "063016",
    year = "2025"
}

@article{DeMarchi:2024zer,
    author = "De Marchi, Andrea Giovanni and Granelli, Alessandro and Nava, Jacopo and Sala, Filippo",
    title = "{Relic Neutrino Background from Cosmic-Ray Reservoirs}",
    eprint = "2405.04568",
    archivePrefix = "arXiv",
    primaryClass = "hep-ph",
    month = "5",
    year = "2024"
}

@article{Blanco:2023dfp,
    author = "Blanco, Carlos and Hooper, Dan and Linden, Tim and Pinetti, Elena",
    title = "{On the Neutrino and Gamma-Ray Emission from NGC 1068}",
    eprint = "2307.03259",
    archivePrefix = "arXiv",
    primaryClass = "astro-ph.HE",
    reportNumber = "FERMILAB-PUB-23-469-T",
    month = "7",
    year = "2023"
}

@article{IceCube:2020acn,
    author = "Aartsen, M. G. and others",
    collaboration = "IceCube",
    title = "{Characteristics of the diffuse astrophysical electron and tau neutrino flux with six years of IceCube high energy cascade data}",
    eprint = "2001.09520",
    archivePrefix = "arXiv",
    primaryClass = "astro-ph.HE",
    doi = "10.1103/PhysRevLett.125.121104",
    journal = "Phys. Rev. Lett.",
    volume = "125",
    number = "12",
    pages = "121104",
    year = "2020"
}

@article{Fang:2025nzg,
    author = "Fang, Ke and Halzen, Francis and Hooper, Dan",
    title = "{Cascaded Gamma-Ray Emission Associated with the KM3NeT Ultrahigh-energy Event KM3-230213A}",
    eprint = "2502.09545",
    archivePrefix = "arXiv",
    primaryClass = "astro-ph.HE",
    doi = "10.3847/2041-8213/adbbec",
    journal = "Astrophys. J. Lett.",
    volume = "982",
    number = "1",
    pages = "L16",
    year = "2025"
}

@article{Wang:2021jic,
    author = "Wang, Jin-Wei and Granelli, Alessandro and Ullio, Piero",
    title = "{Direct Detection Constraints on Blazar-Boosted Dark Matter}",
    eprint = "2111.13644",
    archivePrefix = "arXiv",
    primaryClass = "astro-ph.HE",
    doi = "10.1103/PhysRevLett.128.221104",
    journal = "Phys. Rev. Lett.",
    volume = "128",
    number = "22",
    pages = "221104",
    year = "2022"
}

@article{Gorchtein:2010xa,
    author = "Gorchtein, Mikhail and Profumo, Stefano and Ubaldi, Lorenzo",
    title = "{Probing Dark Matter with AGN Jets}",
    eprint = "1008.2230",
    archivePrefix = "arXiv",
    primaryClass = "astro-ph.HE",
    doi = "10.1103/PhysRevD.82.083514",
    journal = "Phys. Rev. D",
    volume = "82",
    pages = "083514",
    year = "2010",
    note = "[Erratum: Phys.Rev.D 84, 069903 (2011)]"
}

@article{Fiorillo:2023dts,
    author = "Fiorillo, Damiano F. G. and Petropoulou, Maria and Comisso, Luca and Peretti, Enrico and Sironi, Lorenzo",
    title = "{TeV Neutrinos and Hard X-Rays from Relativistic Reconnection in the Corona of NGC 1068}",
    eprint = "2310.18254",
    archivePrefix = "arXiv",
    primaryClass = "astro-ph.HE",
    doi = "10.3847/2041-8213/ad192b",
    journal = "Astrophys. J.",
    volume = "961",
    number = "1",
    pages = "L14",
    year = "2024"
}

@article{Gao:2018mnu,
    author = "Gao, Shan and Fedynitch, Anatoli and Winter, Walter and Pohl, Martin",
    title = "{Modelling the coincident observation of a high-energy neutrino and a bright blazar flare}",
    eprint = "1807.04275",
    archivePrefix = "arXiv",
    primaryClass = "astro-ph.HE",
    doi = "10.1038/s41550-018-0610-1",
    journal = "Nature Astron.",
    volume = "3",
    number = "1",
    pages = "88--92",
    year = "2019"
}

@article{Padovani:2018acg,
    author = "Padovani, P. and Giommi, P. and Resconi, E. and Glauch, T. and Arsioli, B. and Sahakyan, N. and Huber, M.",
    title = "{Dissecting the region around IceCube-170922A: the blazar TXS 0506+056 as the first cosmic neutrino source}",
    eprint = "1807.04461",
    archivePrefix = "arXiv",
    primaryClass = "astro-ph.HE",
    doi = "10.1093/mnras/sty1852",
    journal = "Mon. Not. Roy. Astron. Soc.",
    volume = "480",
    number = "1",
    pages = "192--203",
    year = "2018"
}

@article{Fiorillo:2024akm,
    author = "Fiorillo, Damiano F. G. and Comisso, Luca and Peretti, Enrico and Petropoulou, Maria and Sironi, Lorenzo",
    title = "{A Magnetized Strongly Turbulent Corona as the Source of Neutrinos from NGC 1068}",
    eprint = "2407.01678",
    archivePrefix = "arXiv",
    primaryClass = "astro-ph.HE",
    doi = "10.3847/1538-4357/ad7021",
    journal = "Astrophys. J.",
    volume = "974",
    number = "1",
    pages = "75",
    year = "2024"
}

@article{Greisen,
  title = {End to the Cosmic-Ray Spectrum?},
  author = {Greisen, Kenneth},
  journal = {Phys. Rev. Lett.},
  volume = {16},
  issue = {17},
  pages = {748--750},
  numpages = {0},
  year = {1966},
  month = {Apr},
  publisher = {American Physical Society},
  doi = {10.1103/PhysRevLett.16.748},
  url = {https://link.aps.org/doi/10.1103/PhysRevLett.16.748}
}

@article{Zatsepin:1966jv,
    author = "Zatsepin, G. T. and Kuzmin, V. A.",
    title = "{Upper limit of the spectrum of cosmic rays}",
    journal = "JETP Lett.",
    volume = "4",
    pages = "78--80",
    year = "1966"
}

@article{Padovani:2019xcv,
    author = "Padovani, P. and Oikonomou, F. and Petropoulou, M. and Giommi, P. and Resconi, E.",
    title = "{TXS 0506+056, the first cosmic neutrino source, is not a BL Lac}",
    eprint = "1901.06998",
    archivePrefix = "arXiv",
    primaryClass = "astro-ph.HE",
    doi = "10.1093/mnrasl/slz011",
    journal = "Mon. Not. Roy. Astron. Soc.",
    volume = "484",
    number = "1",
    pages = "L104--L108",
    year = "2019"
}

@article{Cerruti:2018tmc,
    author = "Cerruti, M. and Zech, A. and Boisson, C. and Emery, G. and Inoue, S. and Lenain, J. -P.",
    title = "{Leptohadronic single-zone models for the electromagnetic and neutrino emission of TXS 0506+056}",
    eprint = "1807.04335",
    archivePrefix = "arXiv",
    primaryClass = "astro-ph.HE",
    doi = "10.1093/mnrasl/sly210",
    journal = "Mon. Not. Roy. Astron. Soc.",
    volume = "483",
    number = "1",
    pages = "L12--L16",
    year = "2019",
    note = "[Erratum: Mon.Not.Roy.Astron.Soc. 502, L21--L22 (2021)]"
}

@article{Granelli:2022ysi,
    author = "Granelli, Alessandro and Ullio, Piero and Wang, Jin-Wei",
    title = "{Blazar-boosted dark matter at Super-Kamiokande}",
    eprint = "2202.07598",
    archivePrefix = "arXiv",
    primaryClass = "astro-ph.HE",
    reportNumber = "SISSA 02/2022/FISI",
    doi = "10.1088/1475-7516/2022/07/013",
    journal = "JCAP",
    volume = "07",
    number = "07",
    pages = "013",
    year = "2022"
}

@article{Kotera_2010,
   title={Cosmogenic neutrinos: parameter space and detectabilty from PeV to ZeV},
   volume={2010},
   ISSN={1475-7516},
   url={http://dx.doi.org/10.1088/1475-7516/2010/10/013},
   DOI={10.1088/1475-7516/2010/10/013},
   number={10},
   journal={Journal of Cosmology and Astroparticle Physics},
   publisher={IOP Publishing},
   author={Kotera, K and Allard, D and Olinto, A.V},
   year={2010},
   month=oct, pages={013–013} }

@article{Formaggio:2012cpf,
    author = "Formaggio, J. A. and Zeller, G. P.",
    title = "{From eV to EeV: Neutrino Cross Sections Across Energy Scales}",
    eprint = "1305.7513",
    archivePrefix = "arXiv",
    primaryClass = "hep-ex",
    reportNumber = "FERMILAB-PUB-12-785-E",
    doi = "10.1103/RevModPhys.84.1307",
    journal = "Rev. Mod. Phys.",
    volume = "84",
    pages = "1307--1341",
    year = "2012"
}

@article{Katrin:2024tvg,
    author = "Aker, M. and others",
    collaboration = "Katrin",
    title = "{Direct neutrino-mass measurement based on 259 days of KATRIN data}",
    eprint = "2406.13516",
    archivePrefix = "arXiv",
    primaryClass = "nucl-ex",
    month = "6",
    year = "2024"
}

@book{Giunti:2007ry,
    author = "Giunti, Carlo and Kim, Chung W.",
    title = "{Fundamentals of Neutrino Physics and Astrophysics}",
    doi = "10.1093/acprof:oso/9780198508717.001.0001",
    isbn = "978-0-19-850871-7",
    year = "2007"
}

@article{Dolgov:1997mb,
    author = "Dolgov, A. D. and Hansen, S. H. and Semikoz, D. V.",
    title = "{Nonequilibrium corrections to the spectra of massless neutrinos in the early universe}",
    eprint = "hep-ph/9703315",
    archivePrefix = "arXiv",
    reportNumber = "TAC-1997-010",
    doi = "10.1016/S0550-3213(97)00479-3",
    journal = "Nucl. Phys. B",
    volume = "503",
    pages = "426--444",
    year = "1997"
}

@article{Gandhi:1995tf,
    author = "Gandhi, Raj and Quigg, Chris and Reno, Mary Hall and Sarcevic, Ina",
    title = "{Ultrahigh-energy neutrino interactions}",
    eprint = "hep-ph/9512364",
    archivePrefix = "arXiv",
    reportNumber = "FERMILAB-PUB-95-221-T, CLNS-95-1357, MRI-PHY-16-95, UIOWA-95-06, AZPH-TH-95-15",
    doi = "10.1016/0927-6505(96)00008-4",
    journal = "Astropart. Phys.",
    volume = "5",
    pages = "81--110",
    year = "1996"
}

@ARTICLE{1984MNRAS.207..671B,
       author = {{Beck}, S.~C. and {Beckwith}, S.~V.},
        title = "{Star formation in the nucleus of NGC 253.}",
      journal = {MNRAS},
         year = 1984,
        month = apr,
       volume = {207},
        pages = {671-677},
          doi = {10.1093/mnras/207.4.671},
       adsurl = {https://ui.adsabs.harvard.edu/abs/1984MNRAS.207..671B}
}

@article{Murase:2022dog,
    author = "Murase, Kohta",
    title = "{Hidden Hearts of Neutrino Active Galaxies}",
    eprint = "2211.04460",
    archivePrefix = "arXiv",
    primaryClass = "astro-ph.HE",
    doi = "10.3847/2041-8213/aca53c",
    journal = "Astrophys. J. Lett.",
    volume = "941",
    number = "1",
    pages = "L17",
    year = "2022"
}

@article{Peng:2016nsx,
    author = "Peng, Fang-Kun and Wang, Xiang-Yu and Liu, Ruo-Yu and Tang, Qing-Wen and Wang, Jun-Feng",
    title = "{First detection of GeV emission from an ultraluminous infrared galaxy: Arp 220 as seen with the Fermi Large Area Telescope}",
    eprint = "1603.06355",
    archivePrefix = "arXiv",
    primaryClass = "astro-ph.HE",
    doi = "10.3847/2041-8205/821/2/L20",
    journal = "Astrophys. J. Lett.",
    volume = "821",
    number = "2",
    pages = "L20",
    year = "2016"
}

@article{Yoast-Hull:2013wwa,
    author = "Yoast-Hull, Tova M. and Everett, John E. and Gallagher, J. S. and Zweibel, Ellen G.",
    title = "{Winds, Clumps, and Interacting Cosmic Rays in M82}",
    eprint = "1303.4305",
    archivePrefix = "arXiv",
    primaryClass = "astro-ph.HE",
    doi = "10.1088/0004-637X/768/1/53",
    journal = "Astrophys. J.",
    volume = "768",
    pages = "53",
    year = "2013"
}

@article{Rampadarath:2013vna,
    author = "Rampadarath, Hayden and Morgan, John S. and Lenc, Emil and Tingay, Steven J.",
    title = "{Multi-Epoch Very Long Baseline Interferometric Observations of the Nuclear Starburst Region of NGC 253: Improved modelling of the supernova and star-formation rates}",
    eprint = "1310.8033",
    archivePrefix = "arXiv",
    primaryClass = "astro-ph.CO",
    doi = "10.1088/0004-6256/147/1/5",
    journal = "Astron. J.",
    volume = "147",
    pages = "5",
    year = "2014"
}

@article{Jain:2000pu,
    author = "Jain, P. and McKay, Douglas W. and Panda, S. and Ralston, John P.",
    title = "{Extra dimensions and strong neutrino nucleon interactions above 10**19-eV: Breaking the GZK barrier}",
    eprint = "hep-ph/0001031",
    archivePrefix = "arXiv",
    doi = "10.1016/S0370-2693(00)00647-X",
    journal = "Phys. Lett. B",
    volume = "484",
    pages = "267--274",
    year = "2000"
}

@article{Esteban:2022uuw,
    author = "Esteban, Ivan and Prohira, Steven and Beacom, John F.",
    title = "{Detector requirements for model-independent measurements of ultrahigh energy neutrino cross sections}",
    eprint = "2205.09763",
    archivePrefix = "arXiv",
    primaryClass = "hep-ph",
    doi = "10.1103/PhysRevD.106.023021",
    journal = "Phys. Rev. D",
    volume = "106",
    number = "2",
    pages = "023021",
    year = "2022"
}

@article{Casper:2002sd,
    author = "Casper, D.",
    editor = "Morfin, J. G. and Sakuda, M. and Suzuki, Y.",
    title = "{The Nuance neutrino physics simulation, and the future}",
    eprint = "hep-ph/0208030",
    archivePrefix = "arXiv",
    doi = "10.1016/S0920-5632(02)01756-5",
    journal = "Nucl. Phys. B Proc. Suppl.",
    volume = "112",
    pages = "161--170",
    year = "2002"
}

@article{Eichmann:2022lxh,
    author = {Eichmann, Bj\"orn and Oikonomou, Foteini and Salvatore, Silvia and Dettmar, Ralf-J\"urgen and Becker Tjus, Julia},
    title = "{Solving the Multimessenger Puzzle of the AGN-starburst Composite Galaxy NGC 1068}",
    eprint = "2207.00102",
    archivePrefix = "arXiv",
    primaryClass = "astro-ph.HE",
    doi = "10.3847/1538-4357/ac9588",
    journal = "Astrophys. J.",
    volume = "939",
    number = "1",
    pages = "43",
    year = "2022"
}

@article{Rodrigues:2018tku,
    author = "Rodrigues, Xavier and Gao, Shan and Fedynitch, Anatoli and Palladino, Andrea and Winter, Walter",
    title = "{Leptohadronic Blazar Models Applied to the 2014\textendash{}2015 Flare of TXS 0506+056}",
    eprint = "1812.05939",
    archivePrefix = "arXiv",
    primaryClass = "astro-ph.HE",
    reportNumber = "DESY-19-019",
    doi = "10.3847/2041-8213/ab1267",
    journal = "Astrophys. J. Lett.",
    volume = "874",
    number = "2",
    pages = "L29",
    year = "2019"
}

@article{Hanhart:2001fx,
    author = "Hanhart, Christoph and Pons, Jose A. and Phillips, Daniel R. and Reddy, Sanjay",
    title = "{The Likelihood of GODs' existence: Improving the SN1987a constraint on the size of large compact dimensions}",
    eprint = "astro-ph/0102063",
    archivePrefix = "arXiv",
    doi = "10.1016/S0370-2693(01)00544-5",
    journal = "Phys. Lett. B",
    volume = "509",
    pages = "1--9",
    year = "2001"
}

@article{Hall:1999mk,
    author = "Hall, Lawrence J. and Tucker-Smith, David",
    title = "{Cosmological constraints on theories with large extra dimensions}",
    eprint = "hep-ph/9904267",
    archivePrefix = "arXiv",
    reportNumber = "LBNL-43091, UCB-PTH-99-14, LBL-43091",
    doi = "10.1103/PhysRevD.60.085008",
    journal = "Phys. Rev. D",
    volume = "60",
    pages = "085008",
    year = "1999"
}

@article{CMS:2017zts,
    author = "Sirunyan, A. M. and others",
    collaboration = "CMS",
    title = "{Search for new physics in final states with an energetic jet or a hadronically decaying $W$ or $Z$ boson and transverse momentum imbalance at $\sqrt{s}=13\text{ }\text{ }\mathrm{TeV}$}",
    eprint = "1712.02345",
    archivePrefix = "arXiv",
    primaryClass = "hep-ex",
    reportNumber = "CMS-EXO-16-048, CERN-EP-2017-294",
    doi = "10.1103/PhysRevD.97.092005",
    journal = "Phys. Rev. D",
    volume = "97",
    number = "9",
    pages = "092005",
    year = "2018"
}

@article{IceCube:2025ezc,
    author = "Abbasi, R. and others",
    collaboration = "IceCube",
    title = "{A search for extremely-high-energy neutrinos and first constraints on the ultra-high-energy cosmic-ray proton fraction with IceCube}",
    eprint = "2502.01963",
    archivePrefix = "arXiv",
    primaryClass = "astro-ph.HE",
    month = "2",
    year = "2025"
}

@article{IceCube:2018cha,
    author = "Aartsen, M. G. and others",
    collaboration = "IceCube",
    title = "{Neutrino emission from the direction of the blazar TXS 0506+056 prior to the IceCube-170922A alert}",
    eprint = "1807.08794",
    archivePrefix = "arXiv",
    primaryClass = "astro-ph.HE",
    doi = "10.1126/science.aat2890",
    journal = "Science",
    volume = "361",
    number = "6398",
    pages = "147--151",
    year = "2018"
}

@article{Yuan:2025ctq,
    author = "Yuan, Chengchao and Fiorillo, Damiano F. G. and Petropoulou, Maria and Liu, Qinrui",
    title = "{Coupled Time-Dependent Proton Acceleration and Leptonic-Hadronic Radiation in Turbulent Supermassive Black Hole Coronae}",
    eprint = "2508.08233",
    archivePrefix = "arXiv",
    primaryClass = "astro-ph.HE",
    month = "8",
    year = "2025"
}

@article{KM3NeT:2025ccp,
    author = "Adriani, O. and others",
    collaboration = "KM3NeT",
    title = "{The ultra-high-energy event KM3-230213A within the global neutrino landscape}",
    eprint = "2502.08173",
    archivePrefix = "arXiv",
    primaryClass = "astro-ph.HE",
    month = "2",
    year = "2025"
}

@article{PierreAuger:2015ihf,
    author = "Aab, Alexander and others",
    collaboration = "Pierre Auger",
    title = "{Improved limit to the diffuse flux of ultrahigh energy neutrinos from the Pierre Auger Observatory}",
    eprint = "1504.05397",
    archivePrefix = "arXiv",
    primaryClass = "astro-ph.HE",
    reportNumber = "FERMILAB-PUB-15-150-AD-AE-CD-TD",
    doi = "10.1103/PhysRevD.91.092008",
    journal = "Phys. Rev. D",
    volume = "91",
    number = "9",
    pages = "092008",
    year = "2015"
}

@article{ANITA:2018vwl,
    author = "Gorham, P. W. and others",
    collaboration = "ANITA",
    title = "{Constraints on the diffuse high-energy neutrino flux from the third flight of ANITA}",
    eprint = "1803.02719",
    archivePrefix = "arXiv",
    primaryClass = "astro-ph.HE",
    doi = "10.1103/PhysRevD.98.022001",
    journal = "Phys. Rev. D",
    volume = "98",
    number = "2",
    pages = "022001",
    year = "2018"
}

@article{Stecker:1978ah,
    author = "Stecker, F. W.",
    title = "{Diffuse Fluxes of Cosmic High-Energy Neutrinos}",
    reportNumber = "NASA-TM-79609",
    doi = "10.1086/156919",
    journal = "Astrophys. J.",
    volume = "228",
    pages = "919--927",
    year = "1979"
}

@article{Stecker:1991vm,
    author = "Stecker, F. W. and Done, C. and Salamon, M. H. and Sommers, P.",
    title = "{High-energy neutrinos from active galactic nuclei}",
    reportNumber = "NASA-HEAPTH-91-007",
    doi = "10.1103/PhysRevLett.66.2697",
    journal = "Phys. Rev. Lett.",
    volume = "66",
    pages = "2697--2700",
    year = "1991",
    note = "[Erratum: Phys.Rev.Lett. 69, 2738 (1992)]"
}

@article{Murase:2018utn,
    author = "Murase, Kohta and Fukugita, Masataka",
    title = "{Energetics of High-Energy Cosmic Radiations}",
    eprint = "1806.04194",
    archivePrefix = "arXiv",
    primaryClass = "astro-ph.HE",
    doi = "10.1103/PhysRevD.99.063012",
    journal = "Phys. Rev. D",
    volume = "99",
    number = "6",
    pages = "063012",
    year = "2019"
}

@article{Hooper:2023ssc,
    author = "Hooper, Dan and Plant, Kathryn",
    title = "{Leptonic Model for Neutrino Emission from Active Galactic Nuclei}",
    eprint = "2305.06375",
    archivePrefix = "arXiv",
    primaryClass = "astro-ph.HE",
    reportNumber = "FERMILAB-PUB-23-232-T",
    doi = "10.1103/PhysRevLett.131.231001",
    journal = "Phys. Rev. Lett.",
    volume = "131",
    number = "23",
    pages = "231001",
    year = "2023"
}

@article{Das:2024vug,
    author = "Das, Abhishek and Zhang, B. Theodore and Murase, Kohta",
    title = "{Revealing the Production Mechanism of High-energy Neutrinos from NGC 1068}",
    eprint = "2405.09332",
    archivePrefix = "arXiv",
    primaryClass = "astro-ph.HE",
    doi = "10.3847/1538-4357/ad5a04",
    journal = "Astrophys. J.",
    volume = "972",
    number = "1",
    pages = "44",
    year = "2024"
}

@article{IceCube:2018dnn,
    author = "Aartsen, M. G. and others",
    collaboration = "IceCube, Fermi-LAT, MAGIC, AGILE, ASAS-SN, HAWC, H.E.S.S., INTEGRAL, Kanata, Kiso, Kapteyn, Liverpool Telescope, Subaru, Swift NuSTAR, VERITAS, VLA/17B-403",
    title = "{Multimessenger observations of a flaring blazar coincident with high-energy neutrino IceCube-170922A}",
    eprint = "1807.08816",
    archivePrefix = "arXiv",
    primaryClass = "astro-ph.HE",
    doi = "10.1126/science.aat1378",
    journal = "Science",
    volume = "361",
    number = "6398",
    pages = "eaat1378",
    year = "2018"
}

@article{KM3NeT:2025npi,
    author = "Aiello, S. and others",
    collaboration = "KM3NeT",
    title = "{Observation of an ultra-high-energy cosmic neutrino with KM3NeT}",
    doi = "10.1038/s41586-024-08543-1",
    journal = "Nature",
    volume = "638",
    number = "8050",
    pages = "376--382",
    year = "2025",
    note = "[Erratum: Nature 640, E3 (2025)]"
}

@article{Zhang:2025rqh,
    author = "Zhang, Jiajie and Sandrock, Alexander and Liao, Jiajun and Yue, Baobiao",
    title = "{Impact of coherent scattering on relic neutrinos boosted by cosmic rays}",
    eprint = "2505.04791",
    archivePrefix = "arXiv",
    primaryClass = "hep-ph",
    month = "5",
    year = "2025"
}

@article{Verberne:2021tse,
    author = "Verberne, Sill and Vink, Jacco",
    title = "{The radial supernova remnant distribution in the Galaxy}",
    eprint = "2103.16973",
    archivePrefix = "arXiv",
    primaryClass = "astro-ph.GA",
    doi = "10.1093/mnras/stab940",
    journal = "Mon. Not. Roy. Astron. Soc.",
    volume = "504",
    number = "1",
    pages = "1536--1544",
    year = "2021"
}

@article{Ranasinghe:2022ntj,
    author = "Ranasinghe, S. and Leahy, D.",
    title = "{Distances, Radial Distribution, and Total Number of Galactic Supernova Remnants}",
    eprint = "2209.04570",
    archivePrefix = "arXiv",
    primaryClass = "astro-ph.HE",
    doi = "10.3847/1538-4357/ac940a",
    journal = "Astrophys. J.",
    volume = "940",
    number = "1",
    pages = "63",
    year = "2022"
}

@article{Neronov:2025jfj,
    author = "Neronov, Andrii and Oikonomou, Foteini and Semikoz, Dmitri",
    title = "{KM3-230213A: An Ultra-High Energy Neutrino from a Year-Long Astrophysical Transient}",
    eprint = "2502.12986",
    archivePrefix = "arXiv",
    primaryClass = "astro-ph.HE",
    month = "2",
    year = "2025"
}

@article{Lykken:2007kp,
    author = "Lykken, Joseph and Mena, Olga and Razzaque, Soebur",
    title = "{Ultrahigh-energy neutrino flux as a probe of large extra-dimensions}",
    eprint = "0705.2029",
    archivePrefix = "arXiv",
    primaryClass = "hep-ph",
    reportNumber = "FERMILAB-PUB-07-361-T",
    doi = "10.1088/1475-7516/2007/12/015",
    journal = "JCAP",
    volume = "12",
    pages = "015",
    year = "2007"
}

@article{IceCube-Gen2:2021rkf,
    author = "Abbasi, Rasha and others",
    collaboration = "IceCube-Gen2",
    title = "{Sensitivity studies for the IceCube-Gen2 radio array}",
    eprint = "2107.08910",
    archivePrefix = "arXiv",
    primaryClass = "astro-ph.HE",
    reportNumber = "PoS-ICRC2021-1183",
    doi = "10.22323/1.395.1183",
    journal = "PoS",
    volume = "ICRC2021",
    pages = "1183",
    year = "2021"
}

@article{Van_Den_Bergh_1996, title={Supernova Rates}, volume={145}, DOI={10.1017/S0252921100007855}, journal={International Astronomical Union Colloquium}, author={Van Den Bergh, Sidney}, year={1996}, pages={1–9}}

@article{Meszaros:2001ms,
    author = "Meszaros, Peter and Waxman, Eli",
    title = "{TeV neutrinos from successful and choked gamma-ray bursts}",
    eprint = "astro-ph/0103275",
    archivePrefix = "arXiv",
    doi = "10.1103/PhysRevLett.87.171102",
    journal = "Phys. Rev. Lett.",
    volume = "87",
    pages = "171102",
    year = "2001"
}

@article{Murase:2013ffa,
    author = "Murase, Kohta and Ioka, Kunihito",
    title = "{TeV--PeV Neutrinos from Low-Power Gamma-Ray Burst Jets inside Stars}",
    eprint = "1306.2274",
    archivePrefix = "arXiv",
    primaryClass = "astro-ph.HE",
    doi = "10.1103/PhysRevLett.111.121102",
    journal = "Phys. Rev. Lett.",
    volume = "111",
    pages = "121102",
    year = "2013"
}

@article{Senno:2015tsn,
    author = "Senno, Nicholas and Murase, Kohta and Meszaros, Peter",
    title = "{Choked Jets and Low-Luminosity Gamma-Ray Bursts as Hidden Neutrino Sources}",
    eprint = "1512.08513",
    archivePrefix = "arXiv",
    primaryClass = "astro-ph.HE",
    doi = "10.1103/PhysRevD.93.083003",
    journal = "Phys. Rev. D",
    volume = "93",
    number = "8",
    pages = "083003",
    year = "2016"
}

@article{Bykov:2014lpa,
    author = "Bykov, A. M.",
    title = "{Nonthermal particles and photons in starburst regions and superbubbles}",
    eprint = "1404.0719",
    archivePrefix = "arXiv",
    primaryClass = "astro-ph.HE",
    doi = "10.1007/s00159-014-0077-8",
    journal = "Astron. Astrophys. Rev.",
    volume = "22",
    pages = "77",
    year = "2014"
}

@article{Aharonian:2018oau,
    author = "Aharonian, Felix and Yang, Ruizhi and de O\~na Wilhelmi, Emma",
    title = "{Massive stars as major factories of Galactic cosmic rays}",
    eprint = "1804.02331",
    archivePrefix = "arXiv",
    primaryClass = "astro-ph.HE",
    doi = "10.1038/s41550-019-0724-0",
    journal = "Nature Astron.",
    volume = "3",
    number = "6",
    pages = "561--567",
    year = "2019"
}

@article{Strickland:2009ft,
    author = "Strickland, David K. and Heckman, Timothy M.",
    title = "{Supernova feedback efficiency and mass-loading in the starburst and galactic superwind exemplar M82}",
    eprint = "0903.4175",
    archivePrefix = "arXiv",
    primaryClass = "astro-ph.CO",
    doi = "10.1088/0004-637X/697/2/2030",
    journal = "Astrophys. J.",
    volume = "697",
    pages = "2030--2056",
    year = "2009"
}

@article{HeckmanThompson2017,
    author = "Heckman, Timothy M. and Thompson, Todd A.",
    title = "{Galactic Winds and the Role Played by Massive Stars}",
    eprint = "1701.09062",
    archivePrefix = "arXiv",
    primaryClass = "astro-ph.GA",
    journal = "Handbook of Supernovae",
    year = "2017"
}
\clearpage

\appendix
\section{Comparison with conventional $pp$ and $p\gamma$ neutrino production}
\label{app:ppg_comparison}

\begin{figure}[t!]
\centering
\includegraphics[width=0.48\textwidth]{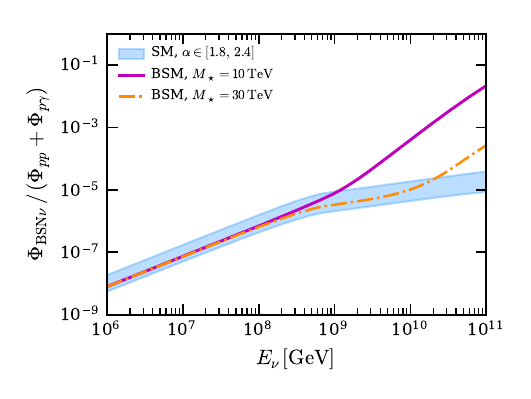}
\caption{Ratio of the cosmic-ray boosted supernova-neutrino flux to the conventional $pp+p\gamma$ neutrino flux at TXS~0506+056, for the same cosmic-ray source model. The blue band brackets the spectral-index range $\alpha\in[1.8,2.4]$ inferred from~\cite{IceCube:2018cha}; the magenta (orange) curve shows the extra-dimensional BSM enhancement for $M_\star=10\,(30)$~TeV. The ratio rises with $E_\nu$ because $\sigma_{p\nu}$ grows with $s$ while $\sigma_{pp}$ is nearly flat.}
\label{fig:ratio_ppg}
\end{figure}

\begin{figure}[H]
\centering
\includegraphics[width=0.48\textwidth]{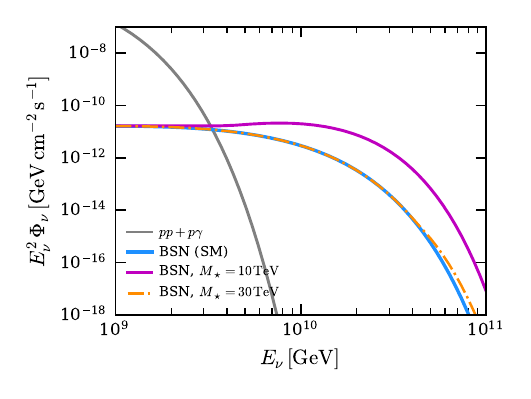}
\caption{Spectral comparison of the cosmic-ray boosted supernova-neutrino flux and the conventional $pp+p\gamma$ neutrino flux at TXS~0506+056 in the transition energy window.}
\label{fig:spectra_BSN}
\end{figure}

In the same sources where supernova neutrinos are boosted by cosmic rays, the same cosmic rays will also produce neutrinos through the conventional $pp$ and $p\gamma$ channels. For a consistent comparison we estimate the boosted component as described in the main text, and the conventional component as

\begin{equation}
\frac{d\Phi_{pp+p\gamma}}{dE_\nu} \simeq (n_p\,\sigma_{pp}+n_\gamma\,\sigma_{p\gamma})\,R_{\rm acc}\,N_\nu\,f_\pi^{-1}\,\Phi_{\rm CR}(E_\nu/f_\pi),
\end{equation}
where $f_\pi\simeq 0.05$ is the average energy fraction carried by each neutrino in $pp$ and $p\gamma$ pion decay. The target-column product $n_p R_{\rm acc}+(\sigma_{p\gamma}/\sigma_{pp})\,n_\gamma R_{\rm acc}$ is fixed so that $\Phi_{pp+p\gamma}$ reproduces the IceCube measurement of TXS~0506+056 at $\sim 100$~TeV for $\alpha=2$. Figure~\ref{fig:ratio_ppg} shows the resulting ratio $\Phi_{\rm BSN\nu}/(\Phi_{pp}+\Phi_{p\gamma})$ as a function of neutrino energy for TXS~0506+056, for the spectral-index range $\alpha\in[1.8,2.4]$ inferred from the time-dependent IceCube analysis~\cite{IceCube:2018cha}, together with representative BSM curves for $M_\star=10$ and $30$~TeV. The ratio grows with energy because $\sigma_{p\nu}$ rises approximately linearly with $s$ regime and then transitions to $\sigma\propto s^{0.363}$ above $\sqrt{s}\sim 4$~TeV~\cite{Formaggio:2012cpf,Gandhi:1995tf}, while $\sigma_{pp}$ is only logarithmically growing in this window.
Two features of the boosted-supernova-neutrino channel may aid its identification. First, the neutrino-proton neutral-current scattering has a much higher inelasticity than pion decay. The boosted neutrino can inherit a sizable fraction of the cosmic-ray energy, $E_\nu^{\max}\sim 0.5 T_p^{\max}$ \cite{Formaggio:2012cpf, Giunti:2007ry}, whereas in the conventional channels $E_\nu\simeq f_\pi\,T_p^{\max}\simeq T_p^{\max}/20$. As a result, for the same cosmic-ray cutoff, the boosted spectrum extends to energies roughly a decade higher than the conventional $pp$ and $p\gamma$ spectra. Second, the different energy dependence of $\sigma_{p\nu}$ relative to $\sigma_{pp}$ and $\sigma_{p\gamma}$ imprints a distinct spectral shape on the boosted component, so that in the ultra-high-energy regime above the conventional cutoff the boosted channel becomes the dominant (or only) neutrino emission from the source. In the presence of the BSM enhancement $\sigma_{p\nu}\propto s^2$ considered in the main text, this feature is accentuated and the ratio can approach unity for $M_\star\lesssim 10$~TeV, reinforcing the complementarity of the two channels and the identification potential at ultra-high energies. These features are made visible in Fig.~\ref{fig:spectra_BSN}, which displays the BSN and conventional $pp+p\gamma$ neutrino spectra at TXS~0506+056 in the energy window where the two channels transition. With a realistic cosmic-ray cutoff near the GZK limit $E_{p,\max}\simeq 5\times 10^{10}$~GeV, the BSM enhancement is concentrated in the upper part of this window: for $M_\star=10$~TeV the BSM curve recovers support from deeper in the suppressed cosmic-ray tail, while for $M_\star=30$~TeV the BSM threshold lies inside that tail and the enhancement is negligible.

\end{document}